\documentclass[aps,prd,draft,showpacs]{revtex4}

\begin{document}

\title{Recursion Relation for the Feynman Diagrams of the Effective Action for the
Third Legendre Transformation}
\author{Chungku Kim}
\date{\today}

\begin{abstract}
We derive a recursion relation of the Feynman diagrams of the
effective action for the third Legendre transformation in case of the
bosonic field theory with cubic interaction. We apply the 
recursion relation to obtain the Feynman diagrams of the effective action
for the third Legendre transformation up to the five-loop order.
The three-particle 
irreducibility of the Feynman diagrams of the effective
action for the third Legendre transformation is shown by induction. 
\end{abstract}
\pacs{11.15.Bt, 12.38.Bx}

\maketitle

\affiliation{Faculty of Mechanical and Automotive Engineering, College of
Engineering, Keimyung University, Daegu 705-701, KOREA}

\section{Introduction}

The effective action plays an important role in studies of the vacuum
instability, dynamical symmetry breaking and the dynamics of composite
particles\cite{Sher} for a given particle physics model. The selective 
resummation of the effective action is important in the investigation of the
equilibrium and the non-equilibrium dynamics of the quantum field theory
\cite{Berges}. The CJT effective action \cite{CJT} which contains only
two-particle-irreducible (2PI) Feynman diagrams \cite{2PI}
was widely used among several resummation schemes. Recently, the nPI effective
action \cite{Berges, nPI} defined by the n-th Legendre transformation
of the generating functional was studied extensively as a generalization
of the CJT effective action which was obtained from the second Legendre 
transformation of the generating functional. Especially, the effective action
for the third Legendre transformation was used in the investigation of the QED
electrical conductivity \cite{Carrington}. However, it was shown that all the
fourth Legendre transformation known yet actually contains the four-particle-reducible
Feynman diagrams \cite{kim1}.
 
The recursive generation of the Feynman diagrams was investigated by using the
functional integral identities $\int D\Phi \ \frac{\delta}
{\delta \Phi }F[\Phi ]=0$ in case of the 
connected and the one-particle-irreducible (1PI) effective action 
was obtained for the multicomponent $\phi ^{4}$-theory, QED and scalar QED theories \cite
{R2,R3,R4,R5,R6,R7,R8}. Recently, we have derived a new method to obtain the recursion 
relation for the ordinary \cite{kim2} as well as CJT effective action \cite{kim3} 
by using the functional derivative identities. 
In this paper, we apply this method to the Feynman diagrams of the effective action for the
third Legendre transformation in case of the bosonic field theory with the cubic interaction.
In Sec.II, we derive the recursion relation 
for the Feynman diagrams of the effective action for the third Legendre
transformation. By using the recursion relation, we obtain the Feynman diagrams up to
the five-loop order.  Then we show the
three-particle-irreducibility of the Feynman diagrams of the effective
action for the third Legendre transformation by induction. 
In Sec.III, we give some discussion and conclusions.


\section{ Recursion Relation for the Feynman Diagrams of the Effective
Action for the Third Legendre Transformation}


In this section, we will first derive a recursion relation for the Feynman
Diagrams of the effective action for the third Legendre transformation for
the bosonic field theory with the cubic interaction. The classical action is
given by 
\begin{equation}
S[\Phi ]= \frac{1}{2} \Phi_{A} D_{0AB}^{-1}\Phi_{B}+\frac{1}{6}g_{ABC}\Phi_{A}
\Phi_{B}\Phi_{C}. \label{1}
\end{equation}
In this paper, we use a notation in which the capital letters contain both
the space-time variables and the internal indices and repeated capital
letters mean both integrations over continuous variables and sums over
internal indices. For example, if the capital letter $A$ contains a
space-time variable $x$ and the internal index $i,$

\begin{equation}
J_{A}\Phi _{A}\equiv \sum_{i}\int d^{4}xJ_{i}(x)\Phi _{i}(x).  \label{2}
\end{equation}
The generating functional $W[J_{1},J_{2},J_{3}]$ is given by

\begin{equation}
W[J_{1},J_{2},J_{3}]=-\hbar \ln \int D\Phi \ Exp[-\frac{1}{\hbar }(S[\Phi
]+J_{1A}\Phi _{A}+\frac{1}{2!}J_{2AB}\Phi _{A}\Phi _{B}+\frac{1}{3!}%
J_{3ABC}\Phi _{A}\Phi _{B}\Phi _{C})],  \label{3}
\end{equation}
where the external source $J_{2AB} $ and $J_{3ABC} $ is symmetric under an
exchange of the indices. The functional derivatives of the $%
W[J_{1},J_{2},J_{3}]$ with respect to the external sources are given by the
classical field $\phi $, the full propagator $G$ and the proper three-vertex 
$V_{3}$ as

\begin{eqnarray}
\frac{\delta W[J_{1},J_{2},J_{3}]}{\delta J_{1A}} &=&\langle \Phi
_{A}\rangle =\phi _{A},  \label{4} \\
2!\frac{\delta W[J_{1},J_{2},J_{3}]}{\delta J_{2AB}} &=&\langle \Phi
_{A}\Phi _{B}\rangle =\phi _{A}\phi _{B}+\hbar G_{AB},  \label{5} \\
3!\frac{\delta W[J_{1},J_{2},J_{3}]}{\delta J_{3ABC}} &=&\langle \Phi
_{A}\Phi _{B}\Phi _{B}\rangle =\phi _{A}\phi _{B}\phi _{C}+\hbar (\phi
_{A}G_{BC}+\phi _{B}G_{AC}+\phi _{C}G_{AB})-\hbar ^{2}G_{AA^{\prime
}}G_{BB^{\prime }}G_{CC^{\prime }}V_{3A^{\prime }B^{\prime }C^{\prime }}.
\label{6}
\end{eqnarray}
From Eqs.(3),(4) ,(5)and (6) we can see that 
\begin{equation}
G_{AB}\equiv -\text{ }\frac{\delta ^{2}W[J_{1},J_{2},J_{3}]}{\delta
J_{1A}\delta J_{1B}}.  \label{7}
\end{equation}
and 
\begin{equation}
G_{AA^{\prime }}G_{BB^{\prime }}G_{CC^{\prime }}V_{3A^{\prime }B^{\prime
}C^{\prime }}\equiv -\text{ }\frac{\delta ^{2}W[J_{1},J_{2},J_{3}]}{\delta
J_{1A}\delta J_{1B}\delta J_{1C}}.  \label{8}
\end{equation}
By inverting Eqs.(4),(5) and (6), one can obtain the functionals $J_{i}[\phi
,G,V_{3}]$ $(i=1,2,3)$. Then, the effective action for the third Legendre
transformation is defined as

\begin{eqnarray}
\Gamma [\phi ,G,V_{3}] &=&W[J_{1},J_{2},J_{3}]-J_{1A}\phi _{A}-\frac{1}{2}%
J_{2AB}(\phi _{A}\phi _{B}+\hbar G_{AB})  \nonumber \\
&&-\frac{1}{6}J_{3ABC}\{\phi _{A}\phi _{B}\phi _{C}+\hbar (\phi
_{A}G_{BC}+\phi _{B}G_{AC}+\phi _{C}G_{AB})-\hbar ^{2}G_{AA^{\prime
}}G_{BB^{\prime }}G_{CC^{\prime }}V_{3A^{\prime }B^{\prime }C^{\prime }}\}.
\label{9}
\end{eqnarray}
From Eqs.(4), (5),(6) and (9), one can obtain the following relations:

\begin{eqnarray}
\frac{\delta \Gamma [\phi ,G,V_{3}]}{\delta \phi _{A}} &=&-J_{1A}-J_{2AB}%
\phi _{B}-\frac{1}{2}J_{3ABC}(\phi _{B}\phi _{C}+\hbar G_{BC}),  \label{10}
\\
\frac{\delta \Gamma [\phi ,G,V_{3}]}{\delta G_{AB}} &=&-\text{ }\frac{\hbar 
}{2}J_{2AB}-\text{ }\frac{\hbar }{2}J_{3ABC}\phi _{C}+\text{ }\frac{\hbar
^{2}}{4}(J_{3ACD}V_{3BC^{\prime }D^{\prime }}+J_{3BCD}V_{3AC^{\prime
}D^{\prime }})G_{CC^{\prime }}G_{DD^{\prime }},  \label{11} \\
\frac{\delta \Gamma [\phi ,G,V_{3}]}{\delta V_{3ABC}} &=&\frac{\hbar ^{2}}{6}%
G_{AA^{\prime }}G_{BB^{\prime }}G_{CC^{\prime }}J_{3A^{\prime }B^{\prime
}C^{\prime }}.  \label{12}
\end{eqnarray}
Also, from Eqs.(3) and (9), we obtain 
\begin{eqnarray}
\exp \{-\frac{1}{\hbar }\Gamma [\phi ,G,V_{3}]\} &=&\int D\Phi \ \text{exp}%
\{-\frac{1}{\hbar }(S(\Phi )+J_{1A}(\Phi _{A}-\phi _{A})+\frac{1}{2}%
J_{2AB}(\Phi _{A}\Phi _{B}-\phi _{A}\phi _{B}-\hbar G_{AB})+\frac{1}{6}%
J_{3ABC}(\Phi _{A}\Phi _{B}\Phi _{C}  \nonumber \\
&&-\phi _{A}\phi _{B}\phi _{C}-\hbar (\phi _{A}G_{BC}+\phi _{B}G_{AC}+\phi
_{C}G_{AB})+\hbar ^{2}G_{AA^{\prime }}G_{BB^{\prime }}G_{CC^{\prime
}}V_{3A^{\prime }B^{\prime }C^{\prime }})\}.  \label{13}
\end{eqnarray}
By expanding the effective action $\Gamma [\phi ,G,V_{3}]$ around $\hbar $,
we can obtain the loop-wise expansion of $\Gamma [\phi ,G,V_{3}]$\cite
{Coleman} as

\begin{equation}
\Gamma [\phi ,G,V_{3}]=\sum_{l=0}\hbar ^{l}\Gamma ^{(l)}[\phi ,G,V_{3}]
\label{14}
\end{equation}
Recently Berges\cite{Berges} has obtained $\Gamma ^{(l)}[\phi ,G,V_{3}]$ up
to three loop order by applying the equivalence hierarchy principle to the
previously known Feynman diagrams for the 2PI effective action; 
\begin{eqnarray}
\Gamma ^{(0)}[\phi ,G] &=&S[\phi ],\text{ }\Gamma ^{(1)}[\phi ,G]=\frac{1}{2}%
Tr\ln G^{-1}-\frac{1}{2}TrG(G^{-1}-D^{-1}),  \label{15} \\
\Gamma ^{(2)}[\phi ,G,J_{3}] &=&\frac{1}{12}%
V_{3ABC}V_{3PQR}G_{AP}G_{BQ}G_{CR}-\frac{1}{6}V_{3ABC}\text{ }%
g_{PQR}G_{AP}G_{BQ}G_{CR}=\frac{1}{12}\begin{picture}(30,20)
\put(15,2){\circle{16}} \put(7,2){\line(1,0){16}} \end{picture} -\frac{1}{6}%
\begin{picture}(40,20) \put(15,2){\circle{16}} \put(7,2){\line(1,0){16}}
\put(25,0){$g$} \end{picture},  \label{16} \\
\Gamma ^{(3)}[\phi ,G,J_{3}] &=&-\frac{1}{24}%
V_{3ABC}V_{3DEF}V_{3PQR}V_{3STU}G_{AD}G_{BP}G_{CS}G_{DQ}G_{ET}G_{RU}=-\frac{1%
}{24}\begin{picture}(40,20) \put(15,3){\circle{20}}
\put(15,13){\line(0,-1){10}} \put(5,3){\line(1,0){20}} \end{picture}
\label{17}
\end{eqnarray}
where 
\begin{equation}
D^{-1}\equiv \frac{\delta ^{2} S[\phi] }{\delta \phi _{A}\delta \phi _{B}}%
=D_{0AB}^{-1}+g_{ABC}\phi _{C},  \label{18}
\end{equation}
and we have used the graphical representation in which a line and a three point
vertex represents the propagator $G$ and the proper three-vertex $V_{3ABC}$ 
respectively and the three point vertex with the letter $g$ means the three point
vertex $g_{ABC}$. In the Appendix, it is shown that $J_{3}[\phi ,G,V_{3}]$ is actually
independent of $\phi $ and hence the Feynman diagrams of the $\Gamma
^{(l)}[\phi ,G,V_{3}](l\geq 2)$ is independent of $\phi $. Then from (12),
we can see that only $\Gamma ^{(0)}$ and $\Gamma ^{(1)}$ can contribute to $%
\frac{\delta \Gamma }{\delta \phi _{C}}$. Then from (15) and (18) we obtain 
\begin{equation}
\frac{\delta ^{2}\Gamma }{\delta \phi _{A}\delta G_{BC}}=\frac{\hbar }{2}%
g_{ABC},  \label{19}
\end{equation}
By taking the derivative $\frac{\delta }{\delta \phi _{C}}$ to (11), we
obtain 
\begin{equation}
\frac{\delta ^{2}\Gamma }{\delta \phi _{C}\delta G_{AB}}=-\text{ }\frac{%
\hbar }{2}\frac{\delta J_{2AB}}{\delta \phi _{C}}-\text{ }\frac{\hbar }{2}%
J_{3ABC}  \label{20}
\end{equation}
By comparing (19) and (20), we obtain 
\begin{equation}
\frac{\delta J_{2AB}}{\delta \phi _{C}}=-J_{3ABC}-g_{ABC}  \label{21}
\end{equation}
Then, since $J_{3}$ is independent of $\phi $, we can write $J_{2}[\phi
,G,V_{3}]$ as 
\begin{equation}
J_{2}[\phi ,G,V_{3}]_{AB}=-(J_{3}[G,V_{3}]_{ABC}+g_{ABC})\phi
_{C}+K_{2}[G,V_{3}]_{AB}  \label{22}
\end{equation}
where $K_{2}$ is the $\phi $ independent part of $J_{2}$. In terms of $%
K_{2}, $ we can write (11) and 
\begin{equation}
\frac{\delta \Gamma [\phi ,G,V_{3}]}{\delta G_{AB}}=-\text{ }\frac{\hbar }{2}%
g_{ABC}\phi _{C}-\text{ }\frac{\hbar }{2}K_{2AB}+\text{ }\frac{\hbar ^{2}}{4}%
(J_{3ACD}V_{3BC^{\prime }D^{\prime }}+J_{3BCD}V_{3AC^{\prime }D^{\prime
}})G_{CC^{\prime }}G_{DD^{\prime }},  \label{23}
\end{equation}
From Eqs.(12),(15) and (16), we obtain 
\begin{equation}
J_{3ABC}^{(0)}=-g_{ABC}+V_{3ABC}\text{, }%
J_{3ABC}^{(1)}=V_{3ADE}V_{3BPQ}V_{3CST}G_{DP}G_{QS}G_{TE}=-%
\begin{picture}(40,20) \put(20,3){\circle{20}} \put(20,17){$A$}
\put(0,1){$B$} \put(34,1){$C$} \put(18,13){\framebox(2,2){}}
\put(9,3){\framebox(2,2){}} \put(29,3){\framebox(2,2){}} \end{picture}
\label{24}
\end{equation}
where a box with an capital letter represents the vertex which have indices
that is not contracted with the propagators attached to it. For example, 
\begin{picture}(70,5) \put(29,-1){\framebox(2,2){}}\put(10,0){\line(1,0){40}}
\put(0,-1){$P$} \put(55,-1){$Q$} \put(25,2){$A$} \end{picture}
means $V_{3AP^{\prime }Q^{\prime }}G_{P^{\prime }P}G_{Q^{\prime }Q}.$ 
From (15),(23) and (24), we obtain 
\begin{equation}
K_{2AB}^{(0)}=G_{AB}^{-1}-D_{0AB}^{-1}\text{, }%
J_{2AB}^{(0)}=-(g_{ABC}+J_{3ABC}^{(0)})\phi _{C}+K_{2AB}^{(0)}=-V_{3ABC}\phi
_{C}+G_{AB}^{-1}-D_{0AB}^{-1}  \label{25}
\end{equation}

Now consider the functional identities satisfied by the two sources $%
J_{1}[\phi ,G,V_{3}]$ and $J_{3}[G,V_{3}]$

\begin{equation}
\frac{\delta J_{1A}}{\delta \phi _{P}}\frac{\delta \phi _{P}}{\delta J_{1B}}+%
\frac{\delta J_{1A}}{\delta G_{PQ}}\frac{\delta G_{PQ}}{\delta J_{1B}}+\frac{%
\delta J_{1A}}{\delta V_{3PQR}}\frac{\delta V_{3PQR}}{\delta J_{1B}}=\delta
_{AB}  \label{26}
\end{equation}
and 
\begin{equation}
\frac{\delta J_{3ACD}}{\delta G_{PQ}}\frac{\delta G_{PQ}}{\delta J_{1B}}+%
\frac{\delta J_{3ACD}}{\delta V_{3PQR}}\frac{\delta V_{3PQR}}{\delta J_{1B}}%
=0  \label{27}
\end{equation}
where we have used the fact that $\frac{\delta J_{3ACD}}{\delta \phi _{P}}=0$%
. By eliminating the term $\frac{\delta V_{3PQRD}}{\delta J_{1B}}$ from
Eqs.(26) and (27), we obtain 
\begin{equation}
\frac{\delta J_{1A}}{\delta \phi _{P}}\frac{\delta \phi _{P}}{\delta J_{1B}}+%
\frac{\delta J_{1A}}{\delta G_{PQ}}\frac{\delta G_{PQ}}{\delta J_{1B}}-\frac{%
\delta J_{1A}}{\delta V_{3PQR}}\Omega _{PQR,CDE}^{-1}\frac{\delta J_{3CDE}}{%
\delta G_{PQ}}\frac{\delta G_{PQ}}{\delta J_{1B}}=\delta _{AB},  \label{28}
\end{equation}
where 
\begin{equation}
\Omega _{ABC,DEF}\equiv \frac{\delta J_{3ABC}}{\delta V_{3DEF}}=\frac{6}{%
\hbar ^{2}}G_{AP}^{-1}G_{BQ}^{-1}G_{APCR}^{-1}\frac{\delta ^{2}\Gamma }{%
\delta V_{3DEF}\delta V_{3PQR}}\text{.}  \label{29}
\end{equation}
From (24) and (29), we obtain 
\begin{equation}
\Omega _{ABC,DEF}^{(0)}=\text{ }\frac{1}{6}(\delta _{AD}\delta _{BE}\delta
_{CR}+\delta _{AE}\delta _{BR}\delta _{CD}+\delta _{AR}\delta _{BD}\delta
_{CE}+\delta _{AD}\delta _{BR}\delta _{CE}+\delta _{AR}\delta _{BE}\delta
_{CD}+\delta _{AE}\delta _{BD}\delta _{CR})\text{.}  \label{30}
\end{equation}
From Eqs.(4),(7) and (8), we obtain

\begin{eqnarray}
\frac{\delta \phi _{A}}{\delta J_{1B}} &=&\frac{\delta ^{2}W[J]}{\delta
J_{C}\delta J_{B}}=-G_{CB},  \label{31} \\
\frac{\delta G_{AB}}{\delta J_{1C}} &=&-\frac{\delta ^{2}W[J]}{\delta
J_{1A}\delta J_{1B}\delta J_{1C}}=G_{AA^{\prime }}G_{BB^{\prime
}}G_{CC^{\prime }}V_{3A^{\prime }B^{\prime }C^{\prime }}  \label{32}
\end{eqnarray}
Also from Eqs.(10),(15),(19) and the fact that $\frac{\delta J_{3ABC}}{%
\delta \phi _{P}}=0$, we obtain 
\begin{eqnarray}
\frac{\delta J_{1A}}{\delta \phi _{P}} &=&-D_{0AB}^{-1}-J_{2AP},  \label{33}
\\
\frac{\delta J_{1A}}{\delta G_{PQ}} &=&-\frac{\hbar }{2}g_{APQ}-\frac{\delta
J_{2AC}}{\delta G_{PQ}}\phi _{C}-\frac{\hbar }{2}J_{3APQ}-\frac{1}{2}(\phi
_{B}\phi _{C}+\hbar G_{BC})\frac{\delta J_{3ABC}}{\delta G_{PQ}},  \label{34}
\\
\frac{\delta J_{1A}}{\delta V_{3PQR}} &=&-\frac{\delta J_{2AC}}{\delta
V_{3PQR}}\phi _{C}-\frac{1}{2}(\phi _{B}\phi _{C}+\hbar G_{BC})\frac{\delta
J_{3ABC}}{\delta V_{3PQR}}.  \label{35}
\end{eqnarray}
By substituting Eqs.(31),(32),(33),(34) and (35) into (28), we obtain 
\begin{eqnarray}
\delta _{AB} &=&(D_{0AP}^{-1}+J_{2AP})G_{PB}+[\{-\frac{\hbar }{2}g_{APQ}-%
\frac{\delta J_{2AC}}{\delta G_{PQ}}\phi _{C}-\frac{\hbar }{2}J_{3APQ}-\frac{%
1}{2}(\phi _{C}\phi _{C^{\prime }}+\hbar G_{CC^{\prime }})\frac{\delta
J_{3ACC^{\prime }}}{\delta G_{PQ}}\}  \nonumber \\
&&+\{\frac{\delta J_{2AC}}{\delta V_{3RST}}\phi _{C}+\frac{1}{2}(\phi
_{C}\phi _{C^{\prime }}+\hbar G_{CC^{\prime }})\frac{\delta J_{3ACC^{\prime
}}}{\delta V_{3RST}}\}\Omega _{RST,DEF}^{-1}\frac{\delta J_{3DEF}}{\delta
G_{PQ}}]\left( G^{3}V_{3}\right) _{BPQ}  \nonumber \\
&=&(D_{0AP}^{-1}+J_{2AP})G_{PB}+[\{-\frac{\hbar }{2}g_{APQ}-\frac{\delta
J_{2AC}}{\delta G_{PQ}}\phi _{C}-\frac{\hbar }{2}J_{3APQ}\}+\phi _{C}\frac{%
\delta J_{2AC}}{\delta V_{3RST}}\Omega _{RST,DEF}^{-1}\frac{\delta J_{3DEF}}{%
\delta G_{PQ}}]\left( G^{3}V_{3}\right) _{BPQ},  \label{36}
\end{eqnarray}
where $\left( G^{3}V_{3}\right) _{BPQ}\equiv G_{BB^{\prime }}G_{PP^{\prime
}}G_{QQ^{\prime }}V_{3B^{\prime }P^{\prime }Q^{\prime }}$ and we have used
(29) to obtain the last line of above equation. Note that in the last line
of the above equation, although there is a terms proportional to $\phi ^{2}$
, they cancel each other ( see (22) ). As a result, one can obtain two
independent equations by extracting terms proportional to $\phi $ and those
independent of $\phi $ respectively. Among these two equations, the one
proportional to  $\phi_{C}$ can be obtained by using (12) and (22) as
\begin{eqnarray}
0 &=&(-g_{APC}-J_{3APC})G_{PB}+[-\frac{\delta K_{2AC}}{\delta G_{PQ}}+\frac{%
\delta K_{2AC}}{\delta V_{3RST}}\Omega _{RST,DEF}^{-1}\frac{\delta J_{3DEF}}{%
\delta G_{PQ}}]\left( G^{3}V_{3}\right) _{BPQ},  \nonumber \\
&=&(-g_{APC}-J_{3APC})G_{PB}+[\frac{2}{\hbar }\frac{\delta ^{2}\Gamma [\phi
,G,V_{3}]}{\delta G_{AC}\delta G_{PQ}}-\hbar
(V_{3APS}J_{3CQT}+_{3AQS}J_{3CPT})G_{ST}  \nonumber \\
&&-\frac{\hbar }{3}G_{RR^{\prime }}G_{SS^{\prime }}G_{TT^{\prime }}\frac{%
\delta J_{3R^{\prime }S^{\prime }T^{\prime }}}{\delta G_{AC}}\Omega
_{RST,DEF}^{-1}\frac{\delta J_{3DEF}}{\delta G_{PQ}}]\left(
G^{3}V_{3}\right) _{BPQ}.  \label{37}
\end{eqnarray}
where we have used (23) to obtain the last line of the above equation.
Then, by multiplying $\frac{\hbar ^{2}}{6}V_{3SBT}G_{AS}G_{CT}$ and by using
(12), we can obtain 
\begin{eqnarray}
V_{3SBT}\frac{\delta \Gamma [\phi ,G,V_{3}]}{\delta V_{3SBT}} &=&\frac{1}{3}[%
\text{ }\hbar \frac{\delta ^{2}\Gamma [\phi ,G,V_{3}]}{\delta G_{AC}\delta
G_{PQ}}-\frac{\hbar ^{3}}{6}G_{RR^{\prime }}G_{SS^{\prime }}G_{TT^{\prime }}%
\frac{\delta J_{3R^{\prime }S^{\prime }T^{\prime }}}{\delta G_{AC}}\Omega
_{RST,DEF}^{-1}\frac{\delta J_{3DEF}}{\delta G_{PQ}}-\frac{\hbar ^{3}}{2}%
(V_{3APS}J_{3CQT}  \nonumber \\
&&+V_{3AQS}J_{3CPT})G_{ST}]\left( G^{2}V_{3}\right) _{ACB}G_{BB^{\prime
}}\left( G^{2}V_{3}\right) _{B^{\prime }PQ}\text{ }-\frac{\hbar ^{2}}{2}%
V_{3ABC}\text{ }g_{PQR}G_{AP}G_{BQ}G_{CR}  \nonumber \\
&=&\frac{1}{3}[\text{ }\hbar \frac{\delta ^{2}\Gamma [\phi ,G,V_{3}]}{\delta
G_{AC}\delta G_{PQ}}-\frac{\hbar ^{3}}{6}G_{RR^{\prime }}G_{SS^{\prime
}}G_{TT^{\prime }}\frac{\delta J_{3R^{\prime }S^{\prime }T^{\prime }}}{%
\delta G_{AC}}\Omega _{RST,DEF}^{-1}\frac{\delta J_{3DEF}}{\delta G_{PQ}}%
]\left( G^{2}V_{3}\right) _{ACB}G_{BB^{\prime }}\left( G^{2}V_{3}\right)
_{B^{\prime }PQ}  \nonumber \\
&&-2\hbar \frac{\delta \Gamma [\phi ,G,V_{3}]}{\delta V_{3CSQ}}%
V_{3APS}G_{AA^{\prime }}V_{3A^{\prime }CB}G_{BB^{\prime }}G_{PP^{\prime
}}V_{3B^{\prime }P^{\prime }Q}-\frac{\hbar ^{2}}{2}V_{3ABC}\text{ }%
g_{PQR}G_{AP}G_{BQ}G_{CR}  \label{38}
\end{eqnarray}
where $\left( G^{2}V_{3}\right) _{ABC}\equiv V_{3AB^{\prime }C^{\prime
}}G_{B^{\prime }B}G_{C^{\prime }C}$. Note that the operation $V_{3SBT}
\frac{\delta \Gamma[\phi ,G,V_{3}]}{\delta V_{3SBT}}$ is equivalent to
multiplying each Feynman diagrams in $\Gamma $ by $N_{3}$ which is the number
of the vertex $V_{3}$. By using the fact that the number of the propagator $L$
is related as $3N_{3}=2L$ and that the Euler formula for the number of the loop
$l$ is given by $l=L-N_{3}+1$, we obtain $N_{3}=2l-2$. Then by using (14), we get
\begin{equation}
  V_{3SBT}\frac{\delta \Gamma[\phi ,G,V_{3}]}{\delta V_{3SBT}}=
\sum_{l=0}(2l-2)\hbar ^{l}\Gamma ^{(l)}[\phi ,G,V_{3}] \label{39}
\end{equation}
Since the last term on the 
right hand side (R.H.S.) of (38) contributes only to $\Gamma ^{(2)}$, the $%
l- $th loop order effective action for the third Legendre transformation 
$\Gamma ^{(l)}[\phi ,G,V_{3}]$ in case of $l\geq 3$ is given by 
\begin{eqnarray}
&&\Gamma ^{(l)}[\phi ,G,V_{3}]=\frac{1}{6(l-1)}[-6\frac{\delta \Gamma
^{(l-1)}[\phi ,G,V_{3}]}{\delta V_{3CSQ}}V_{3APS}G_{AA^{\prime
}}V_{3A^{\prime }CB}G_{BB^{\prime }}G_{PP^{\prime }}V_{3B^{\prime }P^{\prime
}Q} \\
&&+\{\frac{\delta ^{2}\Gamma ^{(l-1)}[\phi ,G,V_{3}]}{\delta G_{AC}\delta
G_{PQ}}-\frac{1}{6}G_{RR^{\prime }}G_{SS^{\prime }}G_{TT^{\prime }}\sum_{%
{{p\geq 1,q\geq 0,r\geq 1}}\atop{{p+q+r=l-3}}}\frac{\delta J_{3R^{\prime
}S^{\prime }T^{\prime }}^{(p)}}{\delta G_{AC}}\Omega _{RST,DEF}^{-1(q)}\frac{%
\delta J_{3DEF}^{(r)}}{\delta G_{PQ}}\}\left( G^{2}V_{3}\right)
_{ACB}G_{BB^{\prime }}\left( G^{2}V_{3}\right) _{B^{\prime }PQ}]  \nonumber
\label{40}
\end{eqnarray}
Equation (40) is the central result of this paper. Each term of this
equation can be obtained as follows;

(i) In order to obtain $6\frac{\delta \Gamma ^{(l-1)}[\phi ,G,V_{3}]}{\delta
V_{3CSQ}}V_{3APS}G_{AA^{\prime }}V_{3A^{\prime }CB}G_{BB^{\prime
}}G_{PP^{\prime }}V_{3B^{\prime }P^{\prime }Q}$ we remove $V_{3CSQ}$ and
replace it with $%
\begin{picture}(40,20) \put(10,0){\line(1,0){20}} 
\put(10,0){\line(1,1){10}} \put(30,0){\line(-1,1){10}} \put(9,-1){\framebox(2,2){}}
\put(29,-1){\framebox(2,2){}} \put(18,10){\framebox(2,2){}}
\put(3,0){$C$}\put(20,12){$S$} \put(32,0){$Q$} \end{picture}$ and then
multiply the results by 6. The result is equivalent to connecting the two
different propagators which are connected to the same three point vertex $%
V_{3}$ of $\Gamma ^{(l-1)}[\phi ,G,V_{3}]$ in all possible way and then
multiply the results by 2.

(ii) In order to obtain $\frac{\delta ^{2}\Gamma ^{(l-1)}[\phi ,G,V_{3}]}{%
\delta G_{AC}\delta G_{PQ}}\left( G^{2}V_{3}\right) _{ACB}G_{BB^{\prime
}}\left( G^{2}V_{3}\right) _{B^{\prime }PQ}$ we remove the two propagators $%
G_{AC}$ and $G_{PQ}$ from $\Gamma ^{(l-1)}$ and replace it with $%
\begin{picture}(45,20) \put(10,10){\line(1,0){20}} \put(10,0){\line(1,0){20}}
 \put(20,0){\line(0,1){10}} \put(3,-2){$A$}
\put(3,10){$P$}\put(32,-2){$C$} \put(32,10){$Q$}\end{picture} .$ The result
is equivalent to connecting the two different propagators of $\Gamma
^{(l-1)}[\phi ,G,V_{3}]$ in all possible way and then multiply the results
by 2.

(iii)In case of the last term of (40), note that $\frac{\delta J_{3DEF}^{(r)}%
}{\delta G_{PQ}}\left( G^{2}V_{3}\right) _{B^{\prime }PQ}$ corresponds to
replacing one of the propagators of $J_{3DEF}^{(q)}$ with 
\begin{picture}(50,5) \put(24,-1){\framebox(2,2){}}\put(5,0){\line(1,0){40}}
\put(20,2){$B'$} \end{picture}. Then, in order to obtain $G_{RR^{\prime
}}G_{SS^{\prime }}G_{TT^{\prime }}\sum_{{{p,q,r}}atop{{p+q+r=l-3}}}\frac{%
\delta J_{3R^{\prime }S^{\prime }T^{\prime }}^{(p)}}{\delta G_{AC}}\Omega
_{RST,DEF}^{-1(q)}\frac{\delta J_{3DEF}^{(r)}}{\delta G_{PQ}}\left(
G^{2}V_{3}\right) _{ACB}G_{BB^{\prime }}\left( G^{2}V_{3}\right) _{B^{\prime
}PQ}$, we replace one of the propagators of $J_{3R^{\prime }S^{\prime
}T^{\prime }}^{(p)}$ with 
\begin{picture}(70,5) \put(29,-1){\framebox(2,2){}}\put(10,0){\line(1,0){40}}
\put(25,2){$B$} \end{picture}
and one of the propagators of $J_{3DEF}^{(q)}$ with 
\begin{picture}(70,5) \put(29,-1){\framebox(2,2){}}\put(10,0){\line(1,0){40}}
\put(25,2){$B'$} \end{picture}
and connect the points $B$ and $B^{\prime }.$ Then connect with $G_{RR^{\prime
}}G_{SS^{\prime }}G_{TT^{\prime }}\Omega _{RST,DEF}^{-1(q)}$. Note that
since $\frac{\delta J_{3ABC}^{(0)}}{\delta G_{PQ}}=0,$ this term contributes
to $\Gamma ^{(l)}[\phi ,G,V_{3}]$ when $l\geq 5.$

Now let us apply (40) to obtain the four and five loop Feynman diagrams of
the effective action for the third order Legendre transformation. From (17)
we obtain 
\begin{eqnarray}
\frac{\delta ^{2}\Gamma ^{(3)}[\phi ,G,V_{3}]}{\delta G_{AC}\delta G_{PQ}}%
\left( G^{2}V_{3}\right) _{ACB}G_{BB^{\prime }}\left( G^{2}V_{3}\right)
_{B^{\prime }PQ} &=&-\frac{1}{4}\begin{picture}(40,40)
\put(5,-10){\line(1,0){28}} \put(5,30){\line(1,0){28}}
\put(5,-10){\line(0,1){40}}\put(13,-10){\line(0,1){40}}
\put(33,-10){\line(0,1){40}} \put(13,10){\line(1,0){20}}
\put(13,-6){\line(2,3){8}} \put(33,26){\line(-2,-3){8}} \end{picture}-%
\begin{picture}(50,40) \put(5,-10){\line(1,0){28}}
\put(13,10){\line(1,0){12}} \put(5,30){\line(1,0){28}}
\put(5,-10){\line(0,1){40}} \put(13,-10){\line(0,1){40}}
\put(25,-10){\line(0,1){40}} \put(33,-10){\line(0,1){40}} \end{picture}
\label{41} \\
6\frac{\delta \Gamma ^{(3)}[\phi ,G,V_{3}]}{\delta V_{3CSQ}}%
V_{3APS}G_{AA^{\prime }}V_{3A^{\prime }CB}G_{BB^{\prime }}G_{PP^{\prime
}}V_{3B^{\prime }P^{\prime }Q} &=&-2\begin{picture}(50,40)
\put(5,-10){\line(1,0){28}} \put(13,10){\line(1,0){12}}
\put(5,30){\line(1,0){28}} \put(5,-10){\line(0,1){40}}
\put(13,-10){\line(0,1){40}} \put(25,-10){\line(0,1){40}}
\put(33,-10){\line(0,1){40}} \end{picture}  \label{42}
\end{eqnarray}
and by substituting these results to (40) we obtain 
\begin{equation}
\Gamma ^{(4)}[\phi ,G,V_{3}]=-\frac{1}{72}\begin{picture}(50,40)
\put(5,-10){\line(1,0){28}} \put(5,30){\line(1,0){28}}
\put(5,-10){\line(0,1){40}}\put(13,-10){\line(0,1){40}}
\put(33,-10){\line(0,1){40}} \put(13,10){\line(1,0){20}}
\put(13,-6){\line(2,3){8}} \put(33,26){\line(-2,-3){8}} \end{picture}
\label{43}
\end{equation}
In case of $l=5$, there is a contribution from the third term of (40) so
that 
\begin{eqnarray}
\frac{\delta ^{2}\Gamma ^{(4)}[\phi ,G,V_{3}]}{\delta G_{AC}\delta G_{PQ}}%
\left( G^{2}V_{3}\right) _{ACB}G_{BB^{\prime }}\left( G^{2}V_{3}\right)
_{B^{\prime }PQ} &=&-\frac{1}{2}\begin{picture}(50,40)
\put(5,-10){\line(1,0){36}} \put(5,30){\line(1,0){36}}
\put(5,-10){\line(0,1){40}}\put(13,-10){\line(0,1){40}}
\put(33,-10){\line(0,1){40}} \put(41,-10){\line(0,1){40}}
\put(13,10){\line(1,0){20}} \put(13,-6){\line(2,3){8}}
\put(33,26){\line(-2,-3){8}} \end{picture} -\frac{1}{2}%
\begin{picture}(50,40) \put(5,-10){\line(1,0){36}}
\put(5,30){\line(1,0){36}} \put(5,-10){\line(0,1){40}}
\put(13,-10){\line(0,1){40}} \put(33,-10){\line(0,1){40}}
\put(41,-10){\line(0,1){40}} \put(13,10){\line(1,0){20}}
\put(13,-6){\line(2,3){8}} \put(30,30){\line(-1,-2){8}} \end{picture}
\label{44} \\
6\frac{\delta \Gamma ^{(4)}[\phi ,G,V_{3}]}{\delta V_{3CSQ}}%
V_{3APS}G_{AA^{\prime }}V_{3A^{\prime }CB}G_{BB^{\prime }}G_{PP^{\prime
}}V_{3B^{\prime }P^{\prime }Q} &=&-\begin{picture}(50,40)
\put(5,-10){\line(1,0){36}} \put(5,30){\line(1,0){36}}
\put(5,-10){\line(0,1){40}} \put(13,-10){\line(0,1){40}}
\put(33,-10){\line(0,1){40}} \put(41,-10){\line(0,1){40}}
\put(13,10){\line(1,0){20}} \put(13,-6){\line(2,3){8}}
\put(30,30){\line(-1,-2){8}} \end{picture}  \label{45} \\
G_{RR^{\prime }}G_{SS^{\prime }}G_{TT^{\prime }}\frac{\delta J_{3R^{\prime
}S^{\prime }T^{\prime }}^{(1)}}{\delta G_{AC}}\Omega _{RST,DEF}^{-1(0)}\frac{%
\delta J_{3DEF}^{(1)}}{\delta G_{PQ}}\left( G^{2}V_{3}\right)
_{ACB}G_{BB^{\prime }}\left( G^{2}V_{3}\right) _{B^{\prime }PQ} &=&3%
\begin{picture}(50,40) \put(5,-10){\line(1,0){36}}
\put(5,30){\line(1,0){36}}
\put(5,-10){\line(0,1){40}}\put(13,-10){\line(0,1){40}}
\put(33,-10){\line(0,1){40}} \put(41,-10){\line(0,1){40}}
\put(13,3){\line(1,0){20}} \put(13,16){\line(1,0){20}} \end{picture} +6%
\begin{picture}(50,40) \put(5,-10){\line(1,0){36}}
\put(5,30){\line(1,0){36}}
\put(5,-10){\line(0,1){40}}\put(13,-10){\line(0,1){40}}
\put(33,-10){\line(0,1){40}} \put(41,-10){\line(0,1){40}}
\put(13,10){\line(1,0){20}} \put(13,-6){\line(2,3){8}}
\put(33,26){\line(-2,-3){8}} \end{picture}  \label{46}
\end{eqnarray}
and by substituting these results to (40) we obtain 
\begin{equation}
\Gamma ^{(5)}[\phi ,G,V_{3}]=-\frac{1}{48}\begin{picture}(50,40)
\put(5,-10){\line(1,0){36}} \put(5,30){\line(1,0){36}}
\put(5,-10){\line(0,1){40}}\put(13,-10){\line(0,1){40}}
\put(33,-10){\line(0,1){40}} \put(41,-10){\line(0,1){40}}
\put(13,3){\line(1,0){20}} \put(13,16){\line(1,0){20}} \end{picture} -\frac{1%
}{16}\begin{picture}(50,40) \put(5,-10){\line(1,0){36}}
\put(5,30){\line(1,0){36}}
\put(5,-10){\line(0,1){40}}\put(13,-10){\line(0,1){40}}
\put(33,-10){\line(0,1){40}} \put(41,-10){\line(0,1){40}}
\put(13,10){\line(1,0){20}} \put(13,-6){\line(2,3){8}}
\put(33,26){\line(-2,-3){8}} \end{picture}  \label{47}
\end{equation}
We can see that the Feynman diagrams of the $\Gamma ^{(4)}[\phi ,G,V_{3}]$
and $\Gamma ^{(5)}[\phi ,G,V_{3}]$ given in (43) and (47) coincide with the
3PI diagrams of the 2PI effective action\cite{Diagram}.

Finally, let us show the three-particle-irreducibility of the Feynman
diagrams of $\Gamma ^{(l)}[\phi ,G,V_{3}]$ by induction. For this
purpose, assume that all the Feynman diagrams of the $\Gamma ^{(k)}[\phi
,G,V_{3}](k<l)$ are 3PI diagrams. Then as discussed in (i) and (ii), the sum
of the first and the second term of R.H.S. of (40) is equal to
connecting the two different propagators which are not connected to the same
vertex $V_{3}$ of $\Gamma ^{(l-1)}[\phi ,G,V_{3}]$ in all possible way.
Since the Feynman diagrams of the $\Gamma ^{(k)}[\phi ,G,V_{3}](k<l)$ are
3PI diagrams by assumption, the sum of the first and the second term of
R.H.S. of (40) gives only 3PI Feynman diagrams ( see (41),(42),(44) and (45)
). Now, by using (30), the perturbative expansion of the $\Omega ^{-1(q)}$
is given by 
\begin{equation}
\Omega _{ABC,DEF}^{-1(q)}=-[\Omega _{ABC,STU}^{(1)}\Omega
_{STU,DEF}^{-1(q-1)}+\Omega _{ABC,STU}^{(2)}\Omega
_{STU,DEF}^{-1(q-2)}+...+\Omega _{ABC,STU}^{(q)}].  \label{48}
\end{equation}
and by repeated use of this equation, $\Omega ^{-1}$ can be obtained as the
sum of the products of $\Omega $'s so that 
\begin{equation}
\Omega _{ABC,DEF}^{-1(q)}=\sum_{{{{q_{1},q_{2},...,q_{n}}}}\atop{{{%
q_{1}+q_{2}+..+.q_{n}{=q}}}}}C_{q_{1}q_{2}...q_{n}}^{(q)}\Omega
_{ABC,A_{1}B_{1}C_{1}}^{(q_{1})}\Omega
_{A_{1}B_{1}C_{1},A_{2}B_{2}C_{2}}^{(q_{2})}...\Omega
_{A_{n-1}B_{n-1}C_{n-1},A_{n}B_{n}C_{n}}^{(q_{n})}  \label{49}
\end{equation}
Hence the third term of (40) is sum of terms of the form 
\begin{equation}
G_{RR^{\prime }}G_{SS^{\prime }}G_{TT^{\prime }}\frac{\delta J_{3R^{\prime
}S^{\prime }T^{\prime }}^{(p)}}{\delta G_{AC}}\Omega
_{RST,R_{1}S_{1}T_{1}}^{(q_{1})}\Omega
_{R_{1}S_{1}T_{1},R_{2}S_{2}T_{2}}^{(q_{2})}...\Omega
_{R_{n-1}S_{n-1}T_{n-1},R_{n}S_{n}T_{n}}^{(q_{n})}\frac{\delta J_{3DEF}^{(r)}%
}{\delta G_{PQ}}\left( G^{2}V_{3}\right) _{ACB}G_{BB^{\prime }}\left(
G^{2}V_{3}\right) _{B^{\prime }PQ}  \label{50}
\end{equation}
with $q_{1}+q_{2}+...+q_{n}=q$. By noting that $\frac{\delta J_{3DEF}^{(r)}}{%
\delta G_{PQ}}\left( G^{2}V_{3}\right) _{B^{\prime }PQ}$ corresponds to
replacing one of the propagators of $J_{3DEF}^{(q)}$ with 
\begin{picture}(50,5) \put(24,-1){\framebox(2,2){}}\put(5,0){\line(1,0){40}}
\put(20,2){$B'$} \end{picture}
, the graphical representation of (50) is given in Fig.1. Note that in
Fig.1, part(a) is three-particle-irreducible since if we connect the points $%
A,B$ and $C$ of $J_{3ABC}^{(k)}$ to the vertex $V_{3ABC}$, we should obtain
the 3PI diagram (see (12)). Similarly, part(b) is three-particle-irreducible
since if we connect the points $A,B$ and $C$ of $\Omega _{ABC,DEF}^{(k)}$
with $%
\begin{picture}(50,20)
\put(10,0){\line(1,0){30}}  \put(25,0){\line(0,1){10}}
\put(0,0){$A$} \put(27,7){$B$} \put(43,0){$C$}
 \end{picture}$ and connect the points $D,E$ and $F$ to the vertex $V_{3DEF}$
, we should obtain the 3PI diagram (see (29)). Other boxes and the oval have
same property. Then, it is clear that Fig.1 is a 3PI diagram.

\section{ Discussions and Conclusions}

In this paper, we have obtained the recursion relation for the effective
action for the third Legendre transformation by using the functional
derivative identity. We have applied the result to the case of Feynman
diagrams up to the five-loop order for the bosonic field theory. Then we
prove the three-particle-irreducibility of the Feynman diagrams of the
effective action for the third Legendre transformation by using the
recursion relation

\begin{acknowledgements}

This research was supported in part by the Institute of Natural Science.
\end{acknowledgements}

\section{Appendix}

In order to see that $J_{3}$ does not depend on $\phi $ , let us consider the
perturbative derivation of $\Gamma ^{(l)}[\phi ,G,V_{3}]$ \cite{kim2} and
define $\overline{\Delta }$ as 
\begin{equation}
\overline{\Delta }[\phi ,G,J_{3}]=W[J_{1},J_{2},J_{3}]-J_{1A}\phi _{A}-\frac{%
1}{2}J_{2AB}(\phi _{A}\phi _{B}+\hbar G_{AB})
\end{equation}
Note that $\overline{\Delta }[\phi ,G,J_{3}]$ is the 2PI effective action 
\cite{CJT} with the classical action given by $\overline{S}[\Phi ]=S[\Phi ]+%
\frac{1}{6}J_{3ABC}\Phi _{A}\Phi _{B}\Phi _{C}$ so that the first three
terms of perturbative expansion of $\overline{\Delta }$ is given by 
\begin{eqnarray}
\overline{\Delta }^{(0)}[\phi ,G,J_{3}] &=&\overline{S}[\phi ]=S[\phi ]+%
\frac{1}{6}J_{3ABC}\phi _{A}\phi _{B}\phi _{C},\text{ }\overline{\Delta }%
^{(1)}[\phi ,G,J_{3}]=\frac{1}{2}Tr\ln G^{-1}-\frac{1}{2}TrG(G^{-1}-%
\overline{D}^{-1}), \\
\text{ }\overline{\Delta }^{(2)}[\phi ,G,J_{3}] &=&-\frac{1}{12}%
(g_{ABC}+J_{3ABC})(g_{PQR}+J_{3PQR})G_{AP}G_{BQ}G_{CR}
\end{eqnarray}
where 
\begin{equation}
\overline{D}_{AB}^{-1}[\phi ]\equiv \frac{\delta ^{2}\overline{S}[\phi ]}{%
\delta \phi _{A}\delta \phi _{B}}=D_{AB}^{-1}[\phi ]+\frac{1}{2}J_{3ABC}\phi
_{C}
\end{equation}
with $D_{AB}^{-1}[\phi ]\equiv \frac{\delta ^{2}S[\phi ]}{\delta \phi
_{A}\delta \phi _{B}}$. The higher order terms $\overline{\Delta }%
^{(k)}(k\succeq 2)$ is composed of 2PI vacuum diagrams with the propagator G
and the three point vertex $g_{ABC}+J_{3ABC}$. It follows that $\overline{%
\Delta }^{(k)}(k\succeq 2)$ does not depend on $\phi $. Next let us define $%
\Delta $ as $_{{}}$%
\begin{equation}
\Delta [\phi ,G,J_{3}]=\overline{\Delta }[\phi ,G,J_{3}]-\frac{1}{6}%
J_{3ABC}[\phi _{A}\phi _{B}\phi _{C}+\hbar (\phi _{A}G_{BC}+\phi
_{B}G_{AC}+\phi _{C}G_{AB})]
\end{equation}
The perturbative expansion of $\Delta $ is given by 
\begin{equation}
\Delta ^{(0)}[\phi ,G,J_{3}]=S[\phi ],\text{ }\Delta ^{(1)}[\phi ,G,J_{3}]=%
\frac{1}{2}Tr\ln G^{-1}-\frac{1}{2}TrG(G^{-1}-D^{-1}),
\end{equation}
and $\Delta ^{(k)}=\overline{\Delta }^{(k)}(k\succeq 2)$ so that $\Delta
^{(k)}(k\succeq 2)$ does not depend on $\phi $. From (3),(9),(51) and (55),
we can see that $\Gamma [\phi ,G,V_{3}]$ can be obtained from $\Delta [\phi
,G,J_{3}]$ by the Legendre transformation with respect to $J_{3}$ as
\begin{equation}
\Gamma [\phi ,G,V_{3}]=\Delta [\phi ,G,J_{3}]+\frac{\hbar ^{2}}{6}%
J_{3ABC}G_{AA^{\prime }}G_{BB^{\prime }}G_{CC^{\prime }}V_{3A^{\prime
}B^{\prime }C^{\prime }}
\end{equation}
In order to determine $J_{3}[\phi ,G,V_{3}]$, let us take the derivative of
(57) with respect to $V_{3}$ as
\begin{equation}
\frac{\delta \Gamma [\phi ,G,V_{3}]}{\delta V_{3PQR}}=(\frac{\delta \Delta
[\phi ,G,J_{3}]}{\delta J_{3ABC}}+\frac{\hbar ^{2}}{6}G_{AA^{\prime
}}G_{BB^{\prime }}G_{CC^{\prime }}V_{3A^{\prime }B^{\prime }C^{\prime }})%
\frac{\delta J_{3ABC}}{\delta V_{3PQR}}+\frac{\hbar ^{2}}{6}%
J_{3ABC}G_{AP}G_{BQ}G_{CR}
\end{equation}
By comparing (58) with (12) we obtain 
\begin{equation}
\frac{\delta \Delta [\phi ,G,J_{3}]}{\delta J_{3ABC}}=-\frac{\hbar ^{2}}{6}%
G_{AA^{\prime }}G_{BB^{\prime }}G_{CC^{\prime }}V_{3A^{\prime }B^{\prime
}C^{\prime }}
\end{equation}
and by using this, we can obtain the perturbative expansion of the $%
J_{3}=\sum_{l=0}\hbar ^{l}J_{3}^{(l)}$ as a functional of the order $\hbar
^{0}$ quantities $\phi ,G$ and $V_{3}$. From (53) and (59), we can see that $%
J_{3ABC}^{(0)}=-g_{ABC}+V_{3ABC}$ which agrees with (24). $J_{3}^{(l)}(l\geq
1)$ can be obtained from $\hbar ^{l+2}$ term of (59). For example, $%
J_{3}^{(1)}$ and $J_{3}^{(2)}$ can be determined from 
\begin{equation}
\left[ \frac{\delta ^{2}\Delta ^{(2)}}{\delta J_{3PQR}\delta J_{3ABC}}%
\right] _{J_{3}=J_{3}^{(0)}}J_{3ABC}^{(1)}+\left[ \frac{\delta \Delta ^{(3)}%
}{\delta J_{3PQR}}\right] _{J_{3}=J_{3}^{(0)}}=0,
\end{equation}
\begin{equation}
\left[ \frac{\delta ^{2}\Delta ^{(2)}}{\delta J_{3PQR}\delta J_{3ABC}}%
\right] _{J_{3}=J_{3}^{(0)}}J_{3ABC}^{(2)}+\left[ \frac{\delta ^{2}\Delta
^{(3)}}{\delta J_{3PQR}\delta J_{3ABC}}\right]
_{J_{3}=J_{3}^{(0)}}J_{3ABC}^{(1)}++\left[ \frac{\delta \Delta ^{(4)}}{%
\delta J_{3PQR}}\right] _{J_{3}=J_{3}^{(0)}}=0.
\end{equation}
and higher orders of $J_{3}^{(l)}(l\geq 3)$ can be obtained by similar
procedure if $J_{3}^{(k)}(k\prec l)$ are determined. Note that the Feynman
diagrams of $\Delta ^{(l)}(l\geq 3)$ consist of the propagator $G$ and the
three-point vertex $J_{3}+g\ $and that $\left[ \frac{\delta \Delta ^{(l)}}{%
\delta J_{3PQR}}\right] _{J_{3}=J_{3}^{(0)}}$ replaces the three-point
vertex of $\frac{\delta \Delta ^{(l)}}{\delta J_{3}}$ by $V_{3}.$ Then,
since $\frac{\delta ^{2}\Delta ^{(2)}}{\delta J_{3PQR}\delta J_{3ABC}}=-%
\frac{1}{36}[$ $G_{AP}G_{BQ}G_{CR}+permutations]$ , the whole procedure to
determine the $J_{3}^{(l)}(l\geq 1)$ does not depend on $\phi $ as long as $%
J_{3ABC}^{(k)}(k\prec l)$ does not depend on $\phi .$ 

\bibliographystyle{plain}
\bibliography{2PI}

\begin{center}
FIGURE CAPTIONS
\end{center}

\begin{picture}(450,180) 
\put(30,97){\oval(30,60)} \put(0,60){\dashbox(50,80)}
\put(100,70){\framebox(30,60)} \put(90,60){\dashbox(50,80)}
\put(15,145){(a)} \put(110,145){(b)}
\put(180,70){\framebox(30,60)} \put(300,70){\framebox(30,60)}
\put(400,97){\oval(30,60)}

\put(30,127){\line(0,1){30}} \put(400,127){\line(0,1){30}} 
\put(30,157){\line(1,0){370}} \put(200,160){$G_{BB^{'}}$} 

\put(45,85){\line(1,0){54}} \put(45,100){\line(1,0){54}} \put(45,115){\line(1,0){54}} 
\put(130,85){\line(1,0){50}} \put(130,100){\line(1,0){50}} \put(130,115){\line(1,0){50}} 
\put(210,85){\line(1,0){30}} \put(210,100){\line(1,0){30}} \put(210,115){\line(1,0){30}} 
\put(250,90){.....} 
\put(270,85){\line(1,0){30}} \put(270,100){\line(1,0){30}} \put(270,115){\line(1,0){30}} 
\put(330,85){\line(1,0){55}} \put(330,100){\line(1,0){55}} \put(330,115){\line(1,0){55}} 

\put(0,30){$\frac{\delta J_{3R^{\prime}S^{\prime }T^{\prime }}^{(p)}}{\delta G_{AC}}
\left( G^{2}V_{3}\right) _{ACB}$} 
\put(60,87){$G_{RR^{\prime }}$} \put(60,102){$G_{SS^{\prime }}$}
\put(60,117){$G_{TT^{\prime }}$}
\put(100,30){$\Omega_{RST,R_{1}S_{1}T_{1}}^{(q_1)}$}
\put(170,30){$\Omega_{R_{1}S_{1}T_{1},R_{2}S_{2}T_{2}}^{(q_2)}$}
\put(260,30){$\Omega_{R_{n-1}S_{n-1}T_{n-1},R_{n}S_{n}T_{n}}^{(q_n)}$} 
\put(370,30){$\frac{\delta J_{3DEF}^{(r)}}{\delta G_{PQ}}
\left(G^{2}V_{3}\right)_{B^{\prime}PQ}$} \end{picture}

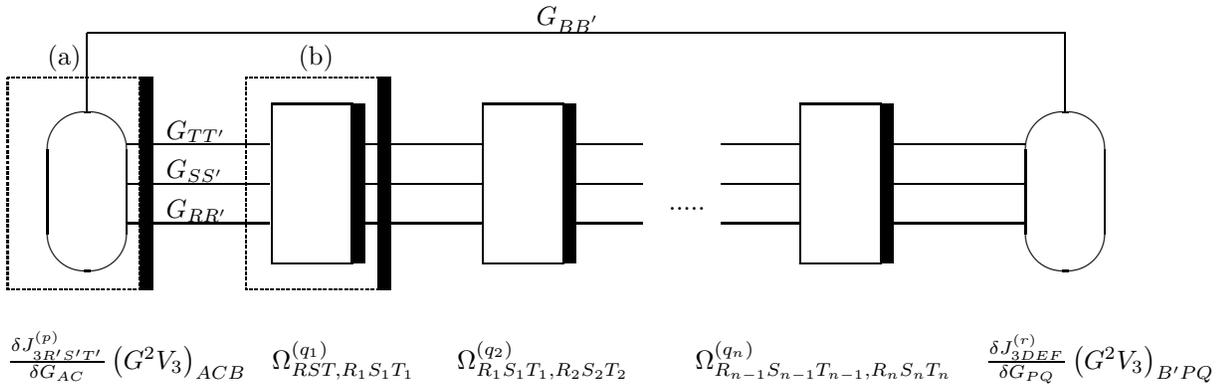
Fig. 1. Graphical representations of the third term of the R.H.S. of Eq.(40).

\end{document}